# Non-Abelian Tensors with Consistent Interactions

Hitoshi NISHINO[1] and Subhash RAJPOOT[2]

*Department of Physics & Astronomy*
*California State University*
*1250 Bellflower Boulevard*
*Long Beach, CA 90840*

## Abstract

We present a systematic method for constructing consistent interactions for a tensor field of an arbitrary rank in the adjoint representation of an arbitrary gauge group in any space-time dimensions. This method is inspired by the dimensional reduction of Scherk-Schwarz, modifying field strengths with certain Chern-Simons forms, together with modified tensorial gauge transformations. In order to define a consistent field strength of a $r$-rank tensor $B_{\mu_1\cdots\mu_r}{}^I$ in the adjoint representation, we need the multiplet $(B_{\mu_1\cdots\mu_r}{}^I,\ B_{\mu_1\cdots\mu_{r-1}}{}^{IJ},\ \cdots,\ B_\mu{}^{I_1\cdots I_r},\ B^{I_1\cdots I_{r+1}})$. The usual problem of consistency of the tensor field equations is circumvented in this formulation.

---

[1] E-Mail: hnishino@csulb.edu
[2] E-Mail: rajpoot@csulb.edu

## 1. Introduction

The problem with coupling a non-Abelian tensor to a non-Abelian gauge field has a long history [1][2][3][4][5][6][7][8]. The question is how to formulate explicit and consistent interactions between a 'physical' tensor carrying an adjoint index and a non-Abelian gauge field. The consistency problem arises already at the classical level, when a covariant divergence is applied to one of the space-time indices in the field equation of such a non-Abelian tensor. This problem is also similar to the so-called 'Velo-Zwanziger disease' encountered for the Rarita-Schwinger field with spin 3/2 or higher-spin fields [9].

The necessity of non-Abelian tensors arises more often [4][8] in the contexts of supergravity [10][11][12]. supersymmetric $\sigma$-models [1], or auxiliary fields for open superstring [6]. However, these works are not so helpful for solving the problem with the minimal coupling of a non-Abelian tensor with a canonical kinetic term to its gauge field. For example, in ref. [1] a non-Abelian tensor field $B_{\mu\nu}{}^I$ was actually introduced. However, this system does not provide the kinetic term for the $B$-field with the minimal coupling to the gauge field $A_\mu{}^I$, because unless the latter is eliminated we can not get the $B$-kinetic term, while this elimination necessarily means the loss of the minimal coupling itself that we wanted. Another proposal for tensor gauge symmetry can be found in [2], but a drawback is that such a symmetry does not commute with Lorentz symmetry. A completely new formulation was proposed with extra spinor fields in [3], that again lacks a canonical kinetic term for the non-Abelian tensor. Also, a more general formulation for non-Abelian tensors was given in [7], but the action invariance imposes additional restrictions on possible non-Abelian gauge groups. In the context of supergravity [11], one solution actually exists [4] based on 'self-duality' condition in odd dimensions [13]. However, this method is valid only in odd space-time dimensions. An invariant field strength for a non-Abelian tensor in supergravity was also presented in ref. [8], but its form is too specific and complicated to be practically useful for more general cases.

We can also try the duality transformation technique [14], but it does not seem to be of much help. Consider in $D$-dimensions the field strength $G_{\mu\nu}{}^I \equiv 2D_{[\mu}B_{\nu]}{}^I \equiv 2(\partial_{[\mu}B_{\nu]}{}^I + f^{IJK}A_{[\mu}{}^J B_{\nu]}{}^K)$ of a vector $B_\mu{}^I$, where $D_\mu$ is the standard non-Abelian covariant derivative with the structure constant $f^{IJK}$. We can try the duality transformation from the 2-nd rank field strength $G_{\mu\nu}{}^I$ into its Hodge dual $H_{\mu_1\cdots\mu_{D-2}}{}^I$ in the adjoint representation that we want. However, the trouble is that the Bianchi identity

$$D_{[\mu}G_{\nu\rho]}{}^I = f^{IJK}F_{[\mu\nu}{}^J B_{\rho]}{}^K \tag{1.1}$$

has the 'bare' potential $B$, preventing such a duality transformation [14]. In view of these problems, the universal formulation of propagating tensors with consistent non-Abelian couplings seems hopelessly difficult to implement.

One clue of defining a consistent field strength for a non-Abelian tensor, however, can be found in the dimensional reduction developed by Scherk and Schwarz [15]. Namely, a field strength singlet in any internal gauge group can produce additional indices in the directions of extra dimensions after the reduction. Accordingly, the field strengths in lower-dimensions have certain extra Chern-Simons (CS) terms playing an important role of canceling unwanted terms arising in the consistency condition of field equations.

Inspired by the dimensional reduction by Scherk-Schwarz [15], we present in this paper the consistent definition of the field strengths of non-Abelian tensor fields of arbitrary ranks, leading to the consistency of their field equations. Such field strengths enable us to construct a large universal class of new consistent interactions.

## 2. Review of the Problem and a Typical Example

We first review the problem with an antisymmetric tensor in the adjoint representation of an arbitrary compact gauge group $G$. Suppose a second-rank tensor field $B_{\mu\nu}{}^I$ in space-time dimension $\forall D$ has the adjoint index $I$ of the group $G$, minimally coupled to a non-Abelian vector field $A_\mu{}^I$. Its naïve field strength is

$$G_{\mu\nu\rho}{}^I \equiv 3D_{[\mu}B_{\nu\rho]}{}^I \equiv 3(\partial_{[\mu}B_{\nu\rho]}{}^I + f^{IJK}A_{[\mu}{}^J B_{\nu\rho]}{}^K) \ , \tag{2.1}$$

where $D_\mu$ is the usual non-Abelian covariant derivative. Now a typical action $I_0 \equiv \int d^4x \, \mathcal{L}_0$ has the lagrangian for the fundamental fields $(B_{\mu\nu}{}^I, A_\mu{}^I)$[3]

$$\mathcal{L}_0 \equiv -\tfrac{1}{12}(G_{\mu\nu\rho}{}^I)^2 - \tfrac{1}{4}(F_{\mu\nu}{}^I)^2 \ , \tag{2.2}$$

yielding their field equations[4]

$$\frac{\delta\mathcal{L}_0}{\delta B_{\mu\nu}{}^I} = +\tfrac{1}{2}D_\rho G^{\mu\nu\rho I} \doteq 0 \ , \tag{2.3a}$$

$$\frac{\delta\mathcal{L}_0}{\delta A_\mu{}^I} = -\tfrac{1}{2}f^{IJK}B_{\nu\rho}{}^J G^{\mu\nu\rho K} - D_\nu F^{\mu\nu I} \doteq 0 \ . \tag{2.3b}$$

The problem arises, when we consider the divergence of the $B$-field equation (2.3a):

$$0 \overset{?}{=} D_\nu\left(\frac{\delta\mathcal{L}_0}{\delta B_{\mu\nu}{}^I}\right) = +\tfrac{1}{4}f^{IJK}F_{\nu\rho}{}^J G^{\mu\nu\rho K} \neq 0 \ , \tag{2.4}$$

---

[3] We use the signature $(-,+,+\cdots,+)$ in this paper.

[4] The symbol $\doteq$ stands for a field equation, to be distinguished from an algebraic identity. We also use the symbol $\overset{?}{=}$ for an equation under question.

unless the field strength $F$ or $G$ vanishes trivially. Since the l.h.s. of (2.4) is supposed to vanish, it leads to an obvious inconsistency already at the classical level.

There is another problem in this system, associated with the tensorial gauge transformation $\delta_\Lambda$ of the $B$-field:

$$\delta_\Lambda B_{\mu\nu}{}^I = +2D_{[\mu}\Lambda_{\nu]}{}^I \ . \tag{2.5}$$

This is because the field strength $G$ is *not* invariant:

$$\delta_\Lambda G_{\mu\nu\rho}{}^I = +3f^{IJK}F_{[\mu\nu}{}^J\Lambda_{\rho]}{}^K \neq 0 \ , \tag{2.6}$$

and therefore the action invariance is lost: $\delta_\Lambda I_0 \neq 0$. These two problems are related to each other, because the non-vanishing of (2.4) is also reformulated as the non-invariance of the action $I_0$ under the tensorial transformation $\delta_\Lambda$.

One way to overcome these problems is to modify the naïve definition (2.1) of the field strength $G$. One clue for such a modification can be found in the dimensional reduction of higher-dimensional tensor fields, originally developed by Scherk and Schwarz [15]. In the dimensional reductions [15], we see that the field strength $\widehat{G}_{\hat{\mu}\hat{\nu}\hat{\rho}\hat{\sigma}}$ [5)] in the original higher-dimensions $D+E$ produces a field strength $G_{\mu\nu\rho\alpha}$ in the final $D$-dimensions, where the index $\alpha$ is in the extra $E$ dimensions. The resulting field strengths have extra CS-terms with the vector field strength $F_{\mu\nu}{}^\alpha$ from the vielbein reduction. To be more specific, $\widehat{G}_{\hat{\mu}\hat{\nu}\hat{\rho}\hat{\sigma}}$ in $D+E$-dimensions yields

$$\widehat{G}_{\hat{\mu}\hat{\nu}\hat{\rho}\hat{\sigma}} \longrightarrow \begin{cases} G_{\mu\nu\rho\sigma} = 4\partial_{[\mu}B_{\nu\rho\sigma]} + 10F_{[\mu\nu}{}^\alpha B_{\rho\sigma]\alpha} \ , \\ G_{\mu\nu\rho\alpha} = 3D_{[\mu}B_{\nu\rho]\alpha} + 6F_{[\mu\nu}{}^\beta B_{\rho]\alpha\beta} \ , \\ G_{\mu\nu\alpha\beta} = 2D_{[\mu}B_{\nu]\alpha\beta} + 2F_{\mu\nu}{}^\gamma B_{\alpha\beta\gamma} - f_{\alpha\beta}{}^\gamma B_{\mu\nu\gamma} \ , \\ G_{\mu\alpha\beta\gamma} = D_\mu B_{\alpha\beta\gamma} + 3f_{[\alpha\beta|}{}^\delta B_{\mu\delta|\gamma]} \ , \\ G_{\alpha\beta\gamma\delta} = -6f_{[\alpha\beta|}{}^\epsilon B_{\epsilon|\gamma\delta]} \ , \end{cases} \tag{2.7}$$

in $D$-dimensions. The $f_{\alpha\beta}{}^\gamma$ is the structure constant for the so-called 'flat group' associated with the Scherk-Schwarz dimensional reduction [15] to give masses to various components in the lower $D$-dimensions. Our prescription here is to mimic this result to the case of a tensor field with non-Abelian group adjoint index, such as $G_{\mu\nu\rho\alpha}$ above. In particular, we replace the extra coordinate indices $\alpha$, $\beta$, $\cdots$ by the adjoint indices $I$, $J$, $\cdots$. We use this as a guide to develop a systematic method to define a consistent field strength for a non-Abelian tensor.

As the first illuminative example, we start with a second-rank tensor $B_{\mu\nu}{}^I$ in the adjoint representation in space-time dimensions $\forall D$. To this end, we need the set of

---

[5)] The fields and indices with *hats* are for the starting $D+E$-dimensional space-time as in [15].

fields $(B_{\mu\nu}{}^I, C_\mu{}^{IJ}, K^{IJK})$ respectively analogous to $B_{\mu\nu\alpha}$, $B_{\mu\alpha\beta}$ and $B_{\alpha\beta\gamma}$ in (2.7), where the indices $I, J, \cdots$ are for the adjoint representation of a compact gauge group $G$, and the indices $IJ$ or $IJK$ are totally antisymmetric. Their field strengths are respectively called $G_{\mu\nu\rho}{}^I$, $H_{\mu\nu}{}^{IJ}$ and $L_\mu{}^{IJK}$. The field strength $G_{\mu\nu\rho}{}^I$ is analogous to $G_{\mu\nu\rho\alpha}$ in (2.7), which has an extra CS term $F_{[\mu\nu|}{}^J C_{|\rho]}{}^{IJ}$. The field strength $H_{\mu\nu}{}^{IJ}$ of $C$ in turn has extra CS-terms $F_{\mu\nu}{}^K K^{IJK}$ and $f^{IJK} B_{\mu\nu}{}^K$ analogous to $G_{\mu\nu\alpha\beta}$ with (2.7). Finally, the field strength $L_\mu{}^{IJK}$ of $K^{IJK}$ has the extra term $\approx f^{[IJ|L} C_\mu{}^{L|K]}$ analogous to $G_{\mu\alpha\beta\gamma}$ in (2.7). Their explicit forms are

$$G_{\mu\nu\rho}{}^I \equiv +3 D_{[\mu} B_{\nu\rho]}{}^I - 3 F_{[\mu\nu|}{}^J C_{|\rho]}{}^{JI} \ , \tag{2.8a}$$

$$H_{\mu\nu}{}^{IJ} \equiv +2 D_{[\mu} C_{\nu]}{}^{IJ} + F_{\mu\nu}{}^K K^{IJK} + f^{IJK} B_{\mu\nu}{}^K \ , \tag{2.8b}$$

$$L_\mu{}^{IJK} \equiv + D_\mu K^{IJK} - 3 f^{[IJ|L} C_\mu{}^{L|K]} \ . \tag{2.8c}$$

For the adjoint indices, we always use superscripts even for contractions, due to the positive definiteness of the compact group $G$. Compared with the naïve definition (2.1), a CS-term is added in (2.8a). Even though there are some discrepancies in the normalizations of the CS-terms in (2.8) compared with (2.7), they are not essential, because the dimensional reduction in [15] is just a 'guide'. Additionally, even though we do not introduce the field strength $G_{\mu\nu\rho\sigma}$, this poses no problem, as we confirm below the tensorial invariance of all the other field strengths. Similarly, we can skip the 'field strength' $G^{IJKL}$ which is an analog of $G_{\alpha\beta\gamma\delta}$ in (2.7), without posing any problem.[6]

All the field strengths in (2.8) are invariant under the tensorial transformations

$$\delta_\Lambda B_{\mu\nu}{}^I = 2 D_{[\mu} \Lambda_{\nu]}{}^I - F_{\mu\nu}{}^J \Lambda^{IJ} \ , \tag{2.9a}$$

$$\delta_\Lambda C_\mu{}^{IJ} = D_\mu \Lambda^{IJ} - f^{IJK} \Lambda_\mu{}^K \ , \tag{2.9b}$$

$$\delta_\Lambda K^{IJK} = 3 f^{[IJ|L} \Lambda^{L|K]} \ , \tag{2.9c}$$

$$\delta_\Lambda A_\mu{}^I = 0 \ . \tag{2.9d}$$

Note that there are two parameters $\Lambda_\mu{}^I$ and $\Lambda^{IJ}$ both carrying non-trivial group indices. By studying the invariance of the field strengths above, we see that it is not enough to have only the tensor field $B_{\mu\nu}{}^I$, but also two other associated fields $C_\mu{}^{IJ}$ and $K^{IJK}$ are needed. In other words, to have an invariant $G_{\mu\nu}{}^I$, we need the total multiplet $(B_{\mu\nu}{}^I, C_\mu{}^{IJ}, K^{IJK})$ with *no* field truncated. Since this is based on the dimensional reductions [15], the number of spacial indices and the adjoint indices always add up to four for all of these field strengths.

---

[6] This sort of 'field strengths' is useful, when we need the potential of scalar fields $K^{IJK}$.

These field strengths also satisfy the Bianchi identities

$$D_{[\mu} G_{\nu\rho\sigma]}{}^I \equiv +\tfrac{3}{2} F_{[\mu\nu}{}^J H_{\rho\sigma]}{}^{IJ} \ , \tag{2.10a}$$

$$D_{[\mu} H_{\nu\rho]}{}^{IJ} \equiv + F_{[\mu\nu}{}^K L_{\rho]}{}^{IJK} + \tfrac{1}{3} f^{IJK} G_{\mu\nu\rho}{}^K \ , \tag{2.10b}$$

$$D_{[\mu} L_{\nu]}{}^{IJK} \equiv + 3 f^{[IJ|L} F_{\mu\nu}{}^M K^{L|KM]} - \tfrac{3}{2} f^{[IJ|L} H_{\mu\nu}{}^{L|K]} \ . \tag{2.10c}$$

Even though the bare 'potential' $K$ is involved in (2.10c), this poses no problem, because $\delta_\Lambda K^{IJK}$ has *no* gradient term. Note that the first term in (2.10c) has all the $IJKM$-indices totally antisymmetrized.

A typical invariant action is

$$\begin{aligned}
I_1 &\equiv \int d^D x \, \mathcal{L}_1 \\
&\equiv \int d^D x \left[ -\tfrac{1}{12}(G_{\mu\nu\rho}{}^I)^2 - \tfrac{1}{4}(H_{\mu\nu}{}^{IJ})^2 - \tfrac{1}{2}(L_\mu{}^{IJK})^2 - \tfrac{1}{4}(F_{\mu\nu}{}^I)^2 \right] \ ,
\end{aligned} \tag{2.11}$$

yielding the field equations

$$\frac{\delta \mathcal{L}_1}{\delta B_{\mu\nu}{}^I} = +\tfrac{1}{2} D_\rho G^{\mu\nu\rho\, I} - \tfrac{1}{2} f^{IJK} H^{\mu\nu\, JK} \doteq 0 \ , \tag{2.12a}$$

$$\frac{\delta \mathcal{L}_1}{\delta C_\mu{}^{IJ}} = -D_\nu H^{\mu\nu\, IJ} + \tfrac{1}{2} F_{\nu\rho}{}^{[I|} G^{\mu\nu\rho|J]} - 3 f^{[I|KL} L^{\mu KL|J]} \doteq 0 \ , \tag{2.12b}$$

$$\frac{\delta \mathcal{L}_1}{\delta K^{IJK}} = +D_\mu L^{\mu\, IJK} - \tfrac{1}{2} F^{\mu\nu\,[I} H_{\mu\nu}{}^{JK]} \doteq 0 \ , \tag{2.12c}$$

$$\begin{aligned}
\frac{\delta \mathcal{L}_1}{\delta A_\mu{}^I} &= -D_\nu F^{\mu\nu\, I} + \tfrac{1}{2} f^{IJK} G^{\mu\rho\sigma\, J} B_{\rho\sigma}{}^K + D_\rho (C_\sigma{}^{IJ} G^{\mu\rho\sigma\, J}) \\
&\quad + 2 f^{IJL} H^{\mu\nu\, JK} C_\nu{}^{LK} - D_\nu (K^{IJK} H^{\mu\nu\, JK}) + 3 f^{IJK} K^{KLM} L^{\mu\, JLM} \doteq 0 \ . \end{aligned} \tag{2.12d}$$

It is not too difficult to show the consistency

$$D_\nu \left( \frac{\delta \mathcal{L}_1}{\delta B_{\mu\nu}{}^I} \right) \doteq 0 \ , \quad D_\mu \left( \frac{\delta \mathcal{L}_1}{\delta C_\mu{}^{IJ}} \right) \doteq 0 \ , \tag{2.13}$$

for the $B$ and $C$-field equations, by the use of $C$ and $B$-field equations themselves, as well as Jacobi identity for the former. In particular, we see how the CS-terms in (2.8) play special roles in canceling all the unwanted terms in these computations, as a solution to the conventional problem (2.4).

We can perform a similar analysis for the $A_\mu{}^I$-field equation, *i.e.*, the conservation of the non-Abelian source current. This computation, however, is more involved. An intermediate

step shows that

$$D_\mu \left(\frac{\delta \mathcal{L}_1}{\delta A_\mu{}^I}\right) = \tfrac{1}{2} f^{IJK}(D_\mu G^{\mu\rho\sigma\,J}) B_{\rho\sigma}{}^K + \tfrac{1}{2} f^{IJK} G^{\mu\rho\sigma\,J} D_\mu B_{\rho\sigma}{}^K + \tfrac{1}{2}[D_\mu, D_\rho](C_\sigma{}^{IJ} G^{\mu\rho\sigma\,J})$$
$$+ \tfrac{1}{2} f^{IJL}(D_\mu H^{\mu\nu\,JK}) C_\nu{}^{LK} + 2 f^{IJL} H^{\mu\nu\,JK} D_\mu C_\nu{}^{LK} - \tfrac{1}{2}[D_\mu, D_\nu](K^{IJK} H^{\mu\nu\,JK})$$
$$+ 3 f^{IJK}(D_\mu K^{KLM}) L^{\mu\,JLM} + 3 f^{IJK} K^{KLM} D_\mu L^{\mu\,JLM} \stackrel{?}{=} 0 \ . \quad (2.14)$$

We can replace the two commutators by the $F$'s, and use field equations for the three divergence terms, while the two rotation terms and one gradient term on $K$ can be replaced by their associated field strengths. Also by the use of Jacobi identity on the structure constants, we see that all the terms in (2.14) cancel themselves, as desired. This conservation is expected, because our total action $I_1$ is invariant also under the non-Abelian symmetry.

We next investigate the mass spectrum of our system (2.11). To this end, we look only into the linear-order terms in the field equations (2.12). First, the linear-order terms in the $B$-field equation (2.12a) can be rewritten as

$$\frac{\delta \mathcal{L}_1}{\delta B_{\mu\nu}{}^I} = +\tfrac{1}{2}\left(D_\rho D^\rho \widehat{B}^{\mu\nu} - a_0 \widehat{B}_{\mu\nu}{}^I\right) + \mathcal{O}(\phi^2) \doteq 0 \ , \quad (2.15)$$

where $\mathcal{O}(\phi^2)$ stands for any bilinear-order terms in fields. The constant $a_0$ is defined by

$$f^{IJK} f^{JKL} = a_0\, \delta^{IL} \quad (a_0 > 0) \ , \quad (2.16)$$

and we have performed the field redefinition

$$\widehat{B}_{\mu\nu}{}^I \equiv B_{\mu\nu}{}^I + 2 a_0^{-1} f^{IJK} D_{[\mu} C_{\nu]}{}^{JK} \ , \quad (2.17)$$

while imposing the $\Lambda_\mu{}^I$-gauge-fixing condition[7]

$$D_\mu \widehat{B}^{\mu\nu\,I} = 0 \ . \quad (2.18)$$

Note that (2.17) can be also regarded as a tensorial $\Lambda_\mu{}^I$-gauge transformation in (2.9a). Eq. (2.15) implies that the $\widehat{B}$-field satisfies the massive Klein-Gordon equation with the non-tachyonic mass[8]

$$m_{\widehat{B}}^2 \equiv a_0 > 0 \ . \quad (2.19)$$

Similarly, the linear-order terms in the $C$-field equation (2.12b) are

$$\frac{\delta \mathcal{L}_1}{\delta C_\mu{}^{IJ}} = +P^{IJ,KL}\left(D_\nu D^\nu \widehat{C}_\mu{}^{KL} - 3 a_0 \widehat{C}_\mu{}^{KL}\right) + \mathcal{O}(\phi^2) \doteq 0 \ , \quad (2.20)$$

---

[7] Notice that this gauge-fixing condition is on $\widehat{B}$, but not the original $B$.

[8] Since our structure constant $f^{IJK}$ carries a mass dimension, the constant $a_0$ has the dimension of (mass)$^2$.

where

$$\widehat{C}_\mu{}^{IJ} \equiv C_\mu{}^{IJ} - a_0^{-1} f^{IJM} f^{MKL} C_\mu{}^{KL} - a_0^{-1} f^{[I|KL} D_\mu K^{KL|J]} \qquad (2.21\text{a})$$

$$= P^{IJ,KL} C_\mu{}^{KL} - a_0^{-1} f^{[I|KL} D_\mu K^{KL|J]} \quad , \qquad (2.21\text{b})$$

and we have imposed the $\Lambda^{IJ}$-gauge-fixing condition

$$D_\mu \widehat{C}^{\mu IJ} = 0 \quad . \qquad (2.22)$$

The $P$'s is a projector is defined by

$$P^{IJ,KL} \equiv \delta^{[I|K} \delta^{|J]L} - a_0^{-1} h^{IJ,KL} \quad , \quad h^{IJ,KL} \equiv f^{IJM} f^{MKL} \quad , \qquad (2.23\text{a})$$

$$P^{IJ,KL} P^{KL,MN} = P^{IJ,MN} \quad , \quad Q^{IJ,KL} Q^{KL,MN} = Q^{IJ,MN} \quad , \quad Q^{IJ,KL} \equiv a_0^{-1} h^{IJ,KL} \quad , \qquad (2.23\text{b})$$

$$P^{IJ,KL} + Q^{IJ,KL} = \delta^{[I|K} \delta^{|J]L} \quad , \quad P^{IJ,KL} Q^{KL,MN} = 0 \quad , \quad Q^{IJ,KL} P^{KL,MN} = 0 \quad . \qquad (3.23\text{c})$$

Eq. (3.23b,c) imply that $P$ and $Q$ are nothing but projectors. Note that the second term in (2.21a) can be regarded as a tensorial $\Lambda_\mu{}^I$-gauge transformation consistent with (2.17), when[9]

$$\Lambda_\mu{}^I \equiv +a_0^{-1} f^{IJK} C_\mu{}^{JK} \quad . \qquad (2.24)$$

The field equation (2.20) implies that only the components of the $\widehat{C}$-field projected out by $P^{IJ,KL}$ satisfy the massive Klein-Gordon equation with the mass

$$m_{\widehat{C}}^2 \equiv 3a_0 > 0 \quad . \qquad (2.25)$$

The dimensionality of the components projected out by the $P$'s coincides with its trace $g(g-3)/2$. This implies that the projectors $P$ and $Q$ respectively project out the original $g(g-1)/2$-dimensional space of the antisymmetric indices $[IJ]$ into $g(g-3)/2$ and $g$-dimensional subspaces. The remaining $g$ components absent in (2.20) are auxiliary fields with no dynamical freedom, and they can be gauged away. Even though this statement is only at the linear-order, we will shortly give an all-order confirmation of this fact.

Finally, the linear-order terms $K$-field equation (2.12c) are simplified into a single term:

$$\frac{\delta \mathcal{L}_1}{\delta K^{IJK}} = D_\mu D^\mu \widehat{K}^{IJK} + \mathcal{O}(\phi^2) \doteq 0 \quad , \qquad (2.26)$$

where

$$\widehat{K}^{IJK} \equiv K^{IJK} - \tfrac{3}{2} a_0^{-1} h^{[IJ|,MN} K^{MN|K]} + \tfrac{3}{2} a_0^{-1} f^{[IJ|L} f^{|K]MN} K^{MNL} \quad . \qquad (2.27)$$

---

[9] The last term in (2.21a) can be regarded as the $\Lambda^{IJ}$-transformation with (2.28) below.

The last two terms can be regarded as a $\Lambda^{IJ}$-tensorial gauge transformation consistent with (2.21) and (2.9b,c), when

$$\Lambda^{IJ} \equiv -a_0^{-1} f^{[I|KL} K^{KL|J]} \quad . \tag{2.28}$$

Eq. (2.26) implies that the $\widehat{K}$-field satisfies the massless Klein-Gordon equation, as desired.

As we have promised, we can confirm that $g$ components $Q^{IJ,KL} C_\mu{}^{KL}$ among the original $g(g-1)/2$ components with respect to the indices $IJ$ in $C_\mu{}^{IJ}$ can be gauged away. For this, we need to prove the existence of an 'extra' symmetry of the action $I_1$. In fact, the extra symmetry

$$\delta_{\text{E}} C_\mu{}^{IJ} \equiv \lambda\, Q^{IJ,KL} C_\mu{}^{KL} \quad , \tag{2.29}$$

with the real parameter $\lambda$ leaves the action $I_1$ invariant: $\delta_{\text{E}} I_1 = 0$. To confirm this, we use the fact that all the field strengths in $\mathcal{L}_1$ are invariant under the tensorial $\Lambda$-transformations, and moreover, the field redefinitions (2.17), (2.21) and (2.27) are nothing but such $\Lambda$-transformations, iff

$$\Lambda_\mu{}^I \equiv + a_0^{-1} f^{IJK} C_\mu{}^{JK} \quad , \tag{2.24}$$

$$\Lambda^{IJ} \equiv - a_0^{-1} f^{[I|KL} K^{KL|J]} \quad . \tag{2.28}$$

Based on this, we can completely replace all the original *un-hatted* fields by *hatted* fields in $\mathcal{L}_1$. We can use this fact to simplify the confirmation of the vanishing of $\delta_{\text{E}} \mathcal{L}_1$, as

$$\begin{aligned} 0 &\stackrel{?}{=} \delta_{\text{E}} \mathcal{L}_1 = (\delta_{\text{E}} C_\mu{}^{IJ}) \frac{\delta \mathcal{L}_1}{\delta_{\text{E}} C_\mu{}^{IJ}} \\ &= (\delta_{\text{E}} C_\mu{}^{IJ}) \left[ \frac{\delta \widehat{B}_{\rho\sigma}{}^K}{\delta C_\mu{}^{IJ}} \frac{\delta \mathcal{L}_1}{\delta \widehat{B}_{\rho\sigma}{}^K} + \frac{\delta \widehat{C}_\rho{}^{KL}}{\delta C_\mu{}^{IJ}} \frac{\delta \mathcal{L}_1}{\delta \widehat{C}_\rho{}^{KL}} + \frac{\delta \widehat{K}^{KLM}}{\delta C_\mu{}^{IJ}} \frac{\delta \mathcal{L}_1}{\delta \widehat{K}^{KLM}} \right] \\ &= + \tfrac{1}{2} a_0^{-1} \lambda f^{IJK} C_\rho{}^{IJ} f^{KLM} F_{\sigma\tau}{}^L \widehat{G}^{\rho\sigma\tau\, M} \\ &\quad - a_0^{-1} \lambda f^{IJK} f^{KMN} C_\rho{}^{MN} D_\sigma \widehat{H}^{\rho\sigma\, IJ} \quad . \end{aligned} \tag{2.30}$$

There is no contribution from $\delta \widehat{K}/\delta C$. We can use the $\widehat{C}$-field equation (*i.e.*, (2.12b) with all the fields replaced by *hatted* fields) for the last line, and show that a term arising in there cancels the penultimate line of (2.30), while the rest vanishes by itself by Jacobi identity, yielding $\delta_{\text{E}} I_1 = 0$, as desired. This implies that the action $I_1$ is really invariant under the extra symmetry (2.29) to all-orders, and therefore, the $g$ components $Q^{IJ,KL} C_\mu{}^{KL}$ can be completely gauged away.

Note that the proof above is based on the extra symmetry (2.29) valid to all orders. The importance of this all-order confirmation is elucidated as follows: Even though these unphysical components are absent at the linear order in the $C$-field equation (2.20) or at the bilinear order in the lagrangian $\mathcal{L}_1$, they might still enter higher-order terms, generating undesirable constraint equations that complicate our system. Thanks to the all-order proof above, we can safely conclude that those $g$ components are really unphysical and gauged away by the extra symmetry $\delta_{\rm E}$ without spoiling the interactions of other physical fields.

As we have seen so far, our system (2.11) has desirable features of physical system in $\forall D$-dimensional space-time, such as non-tachyonic massive propagating fields, and certain components are non-propagating and are completely gauged away. This analysis provides an additional support to our lagrangian (2.11) as a consistent physical system ready for practical applications.

## 3. Generalizations to Higher-Rank Non-Abelian Tensors

The example in the previous section is only for the 2nd-rank potential $B_{\mu\nu}{}^I$ in the adjoint representation. However, once we have understood the pattern, we can generalize this to more general higher-rank tensor potentials in $\forall D$. Here we require that the highest-rank tensor field $B_{\mu_1\cdots\mu_r}$ to be in the adjoint representation of the gauge group $G$.

In order to simplify the notation, we use the language of differential forms from now on. We omit the usual wedge symbol $\wedge$ for multiplications of forms in order to save space. We also normalize the products, e.g., $AB \equiv A \wedge B \equiv (1/2!) A_\mu dx^\mu \wedge B_\nu dx^\nu$. The $r$-form potential is denoted by $B_r{}^I$ with the subscript $r$ showing its rank, while its field strength $G_{r+1}{}^I$ has an extra CS-term that requires the existence of the $(r-1)$-rank potential $B_{r-1}{}^{IJ}$, which in turn requires the $(r-2)$-rank potential $B_{r-2}{}^{IJK}$, and so forth. After all, we need the set of the $B$-potential fields $(B_r{}^I, B_{r-1}{}^{IJ}, B_{r-2}{}^{IJK}, \cdots, B_0{}^{I_1\cdots I_{r+1}})$ in addition to the non-Abelian gauge field $A_\mu{}^I$. Accordingly, the field strengths are defined for general integer $r \geq 2$ by

$$G_{r+1}{}^I \equiv (r+1)DB_r{}^I - \tfrac{r+1}{r-1} F^J B_{r-1}{}^{JI} , \tag{3.1a}$$

$$G_r{}^{IJ} \equiv rDB_{r-1}{}^{IJ} - \tfrac{r}{r-2} F^J B_{r-2}{}^{KIJ} + \tfrac{1}{2}r(r-1)f^{IJK} B_r{}^K , \tag{3.1b}$$

$$G_{r-1}{}^{IJK} \equiv (r-1)DB_{r-2}{}^{IJK} - \tfrac{r-1}{r-3} F^L B_{r-3}{}^{LIJK} + \tfrac{3}{2}(r-1)(r-2) f^{[IJ|M} B_{r-1}{}^{M|K]} , \tag{3.1c}$$

............................................................

$$\begin{aligned} G_{n+1}{}^{I_1\cdots I_{r-n+1}} \equiv &(n+1)DB_n{}^{I_1\cdots I_{r-n+1}} - \tfrac{n+1}{n-1} F^J B_{n-1}{}^{JI_1\cdots I_{r-n+1}} \\ &+ \tfrac{1}{4}n(n+1)(r-n)(r-n+1) f^{[I_1 I_2|J} B_2{}^{J|I_3\cdots I_r]} \quad (2 \leq n \leq r-1) , \end{aligned} \tag{3.1d}$$

............................................................

$$G_2{}^{I_1\cdots I_r} \equiv 2DB_1{}^{I_1\cdots I_r} - \frac{r(r+1)}{2d_0} F^K B_0{}^{KI_1\cdots I_r} + \tfrac{1}{2}r(r-1)f^{[I_1 I_2|J} B_2{}^{J|I_3\cdots I_r]} \;, \tag{3.1e}$$

$$G_1{}^{I_1\cdots I_{r+1}} \equiv DB_0{}^{I_1\cdots I_{r+1}} + d_0 f^{[I_1 I_2|J} B_1{}^{J|I_3\cdots I_{r+1}]} \;, \tag{3.1f}$$

where $D$ is the covariant derivative form: $DX \equiv dX + [A, X]$,[10] while $d_0$ in (3.1e,f) is an arbitrary non-zero real constant, which is *a priori* arbitrary, but can be fixed to be a non-zero constant.

These field strengths have been fixed in such a way that they are invariant under the following set of tensorial transformations[11]

$$\delta_\Lambda B_r{}^I = rD\Lambda_{r-1}{}^I + F^J \Lambda_{r-2}{}^{JI} \;, \tag{3.2a}$$

$$\delta_\Lambda B_{r-1}{}^{IJ} = (r-1)D\Lambda_{r-2}{}^{IJ} + F^K \Lambda_{r-3}{}^{KIJ} - \tfrac{1}{2}r(r-1)f^{IJK}\Lambda_{r-1}{}^K \;, \tag{3.2b}$$

$$\delta_\Lambda B_{r-2}{}^{IJK} = (r-2)D\Lambda_{r-3}{}^{IJK} + F^L \Lambda_{r-4}{}^{LIJK} - \tfrac{3}{2}(r-1)(r-2)f^{[IJ|L}\Lambda_{r-2}{}^{L|K]} \;, \tag{3.2c}$$

$$\cdots\cdots\cdots\cdots\cdots\cdots\cdots\cdots\cdots\cdots\cdots\cdots\cdots\cdots\cdots\cdots\cdots\cdots\cdots\cdots\cdots$$

$$\delta_\Lambda B_n{}^{I_1\cdots I_{r-n+1}} = nD\Lambda_{n-1}{}^{I_1\cdots I_{r-n+1}} + F^J \Lambda_{n-2}{}^{JI_1\cdots I_{r-n+1}}$$
$$- \tfrac{1}{4}n(n+1)(r-n)(r-n+1)f^{[I_1 I_2|J}\Lambda_n{}^{J|I_3\cdots I_{r-n+1}]} \quad (2 \le n \le r-1) \;, \tag{3.2d}$$

$$\cdots\cdots\cdots\cdots\cdots\cdots\cdots\cdots\cdots\cdots\cdots\cdots\cdots\cdots\cdots\cdots\cdots\cdots\cdots\cdots\cdots$$

$$\delta_\Lambda B_2{}^{I_1\cdots I_{r-1}} = 2D\Lambda_1{}^{I_1\cdots I_{r-1}} + F^J \Lambda_0{}^{JI_1\cdots I_{r-1}} - \tfrac{3}{2}(r-1)(r-2)f^{[I_1 I_2|J}\Lambda_2{}^{J|I_3\cdots I_{r-1}]} \;, \tag{3.2e}$$

$$\delta_\Lambda B_1{}^{I_1\cdots I_r} = D\Lambda_0{}^{I_1\cdots I_r} - \tfrac{1}{2}r(r-1)f^{[I_1 I_2|J}\Lambda_1{}^{J|I_3\cdots I_r]} \;, \tag{3.2f}$$

$$\delta_\Lambda B_0{}^{I_1\cdots I_{r+1}} = -d_0 f^{[I_1 I_2|J}\Lambda_0{}^{J|I_3\cdots I_{r+1}]} \;. \tag{3.2g}$$

The coefficients of these terms have been fixed also by the requirement of the invariance of all the field strengths:

$$\delta_\Lambda G_n{}^{I_1\cdots I_{r-n+2}} = 0 \quad (1 \le n \le r+1) \;. \tag{3.3}$$

The only input we need is that all the coefficients of the $F\Lambda$-terms in (3.2) are normalized to unity, but they are not essential, as long as they are non-zero constants. All the field strengths are also manifestly covariant under the usual non-Abelian gauge transformations $\delta_\alpha$:

$$\delta_\alpha B_n{}^{I_1\cdots I_{r-n+1}} = -(r-n+1)f^{[I_1|JK}\alpha^J B_n{}^{K|I_2\cdots I_{r-n+1}]} \;, \tag{3.4a}$$

---

[10] This operator $D$ should not be confused with the space-time dimension $D$. Such a confusion can be prevented by keeping track of the rank of each term.

[11] As in the case of the $B$'s, the subscripts such as $r-1$ on the $\Lambda$'s are for their ranks.

$$\delta_\alpha A_\mu{}^I = D_\mu \alpha^I ~, \tag{3.4b}$$

$$\delta_\alpha G_{n+1}{}^{I_1\cdots I_{r-n+1}} = -(r-n+1) f^{[I_1|JK} \alpha^J G_{n+1}{}^{K|I_2\cdots I_{r-n+1}]} \qquad (0 \le n \le r) ~. \tag{3.4c}$$

These field strengths also satisfy their proper Bianchi identities:

$$DG_{n+1}{}^{I_1\cdots I_{r-n+1}} \equiv + \frac{n(n+1)(r-n)(r-n+1)}{4(n+2)} f^{[I_1 I_2|J} G_{n+2}{}^{J|I_3\cdots I_{r-n+1}]}$$
$$- \frac{n+1}{n(n-1)} F^J G_n{}^{JI_1\cdots I_{r-n+1}} \qquad (2 \le n \le r) ~, \tag{3.5a}$$

$$DG_2{}^{I_1\cdots I_r} \equiv + \tfrac{1}{6} r(r-1) f^{[I_1 I_2|J} G_3{}^{J|I_3\cdots I_r]} ~, \tag{3.5b}$$

$$DG_1{}^{I_1\cdots I_{r+1}} \equiv - \tfrac{1}{4}(r+1)(r+2) f^{[I_1 I_2|K} F^J B_0{}^{K|JI_3\cdots I_{r+1}]}$$
$$+ \tfrac{1}{2} d_0 f^{[I_1 I_2|J} G_2{}^{J|I_3\cdots I_{r+1}]} ~. \tag{3.5c}$$

The example in section 2 is now just a special case of $r=2,\ n=2$ in (3.1) through (3.5).

Based on these field strengths, we can easily construct non-trivial interacting actions, such as

$$I_2 \equiv \int d^D x\, \mathcal{L}_2$$
$$\equiv \int d^D x \Big[ -\tfrac{1}{r! \cdot 2} (G_{\mu_1\cdots \mu_r}{}^I)^2 - \tfrac{1}{(r-1)!\cdot 2} (G_{\mu_1\cdots \mu_{r-1}}{}^{IJ})^2 - \cdots - \tfrac{1}{(n+1)!\cdot 2} (G_{\mu_1\cdots \mu_{n+1}}{}^{I_1\cdots I_{r-n+1}})^2$$
$$- \cdots - \tfrac{1}{4} (G_{\mu\nu}{}^{I_1\cdots I_r})^2 - \tfrac{1}{2}(G_\mu{}^{I_1\cdots I_{r+1}})^2 - \tfrac{1}{4}(F_{\mu\nu}{}^I)^2 \Big] ~. \tag{3.6}$$

Needless to say, this lagrangian has an enormous number of non-trivial and consistent interactions among these higher-rank tensor fields.

In fact, we can confirm in a way similar to the previous section the consistency of the $B$-field equations of all the fields in this system, which are

$$\frac{\delta \mathcal{L}_2}{\delta B_{\mu_1\cdots \mu_n}{}^{I_1\cdots I_{r-n+1}}} = + \frac{1}{n!} D_\nu G^{\nu\mu_1\cdots \mu_n\, I_1\cdots I_{r-n+1}} - \frac{(r-n+1)(r-n+2)}{4(n-2)!} f^{JK[I_1|} G^{\mu_1\cdots \mu_n\, JK|I_2\cdots I_{r-n+1}]}$$
$$+ \frac{1}{n(n+1)!} F_{\rho\sigma}{}^{[I_1|} G^{\rho\sigma\mu_1\cdots \mu_n|I_2\cdots I_{n-r+1}]} \doteq 0 \qquad (2 \le n \le r) ~, \tag{3.7a}$$

$$\frac{\delta \mathcal{L}_2}{\delta B_\mu{}^{I_1\cdots I_r}} = + D_\nu G^{\nu\mu\, I_1\cdots I_r} - d_0 f^{JK[I_1|} G^{\mu\, JK|I_2\cdots I_r]} + \tfrac{1}{2} F_{\rho\sigma}{}^{[I_1|} G^{\rho\sigma\mu|I_2\cdots I_r]} \doteq 0 ~, \tag{3.7b}$$

$$\frac{\delta \mathcal{L}_2}{\delta B^{I_1\cdots I_{r+1}}} = + D_\mu G^{\mu\, I_1\cdots I_{r+1}} + \frac{r(r+1)}{4 d_0} F_{\mu\nu}{}^{[I_1|} G^{\mu\nu|I_2\cdots I_r]} \doteq 0 ~. \tag{3.7c}$$

For the special case of $n=r$ in (3.7a), its last term does not exist. The consistency of $B$-field equations are the reflection of the invariance $\delta_\Lambda I_2 = 0$, while that of the $A_\mu$-field equation is the non-Abelian invariance $\delta_\alpha I_2 = 0$.

## 5. Concluding Remarks

In this paper, we have presented an explicit, systematic and straightforward method of constructing the consistent field strength for a $r$-th rank tensor $B_r{}^I$ in the adjoint representation. The appropriate field strength $G_{r+1}{}^I$ needs an extra CS-form that needs an $r$-th rank tensor $B_r{}^{IJ}$, whose filed strength in turn needs an $(r-1)$-th rank tensor $B_{r-1}{}^{IJK}$, and this chain continues down to $B_0{}^{I_1\cdots I_{r+1}}$. These field strengths are not only invariant under tensorial transformations $\delta_\Lambda$, and covariant under the gauge transformation $\delta_\alpha$, but they also satisfy their appropriate Bianchi identities (3.3) containing only the field strengths, except for $G_1{}^{I_1\cdots I_{r+1}}$. The last feature poses no problem, because the 'potential' $B_0{}^{I_1\cdots I_{r+1}}$ does not have any gradient of the parameter $\Lambda$ in $\delta_\Lambda B_0{}^{I_1\cdots I_{r+1}}$.

Our method is inspired by the simple dimensional reduction by Scherk and Schwarz in the 1970's [15]. Therefore, there are some features shared with the dimensional reduction in [15], such as the inevitability of introducing the lower-rank tensors $B_{r-1}{}^{IJ}, \cdots, B_0{}^{I_1\cdots I_{r+1}}$, when we start with the highest one $B_r{}^I$, in order to keep the invariance $\delta_\Lambda G_{r+1}{}^I = 0$. Also, all the fields in our multiplet $(B_r{}^I, B_{r-1}{}^{IJ}, B_{r-2}{}^{IJK}, \cdots, B_1{}^{I_1\cdots I_r}, B_0{}^{I_1\cdots I_{r+1}})$ share the common feature that the sum of the rank and the number of the adjoint indices is always $r+1$.

As already stated, all the fields $(B_r{}^I, B_{r-1}{}^{IJ}, B_{r-2}{}^{IJK}, \cdots, B_1{}^{I_1\cdots I_r}, B_0{}^{I_1\cdots I_{r+1}})$ are required for defining the consistent field strength for $B_r{}^I$. In this sense, we can regard them as a 'multiplet' in which all the fields are required. For example, we easily see in the $r=2$ multiplet $(B_{\mu\nu}{}^I, C_\mu{}^{IJK}, K^{IJK})$ that the truncation of the field $K^{IJK}$ leads to $H_{\mu\nu}{}^{IJ}$ without the $FK$-term in (2.8b), lacking the counter term $\approx f^{IJK}F_{\mu\nu}{}^L\Lambda^{KL}$ in $\delta_\Lambda H_{\mu\nu}{}^{IJ}$, and spoiling the invariance: $\delta_\Lambda H_{\mu\nu}{}^{IJ} \neq 0$. Similarly, the truncation of $B_{\mu\nu}{}^I$ leads to $H_{\mu\nu}{}^{IJ}$ without the $f^{IJK}B_{\mu\nu}{}^K$-term, lacking the $f^{IJK}F_{\mu\nu}{}^L\Lambda^{KL}$-term in $\delta_\Lambda H_{\mu\nu}{}^{IJ}$, and destroying again the invariance of $H$.

Compared with the definitions of field strengths in the context of supergravity, *e.g.*, in [8], our field strengths (2.8) or (3.1) are much simpler, and our method is more systematically applicable to a non-Abelian tensor of an arbitrary rank in any space-time dimensions with any gauge groups.

Once we have succeeded in defining the $\delta_\Lambda$-invariant field strengths of higher-rank tensors in the adjoint representations, we can easily construct lagrangians including their kinetic terms. Note that any such lagrangian has non-trivial interaction terms, in addition to their free kinetic terms. Since all of these field strengths are invariant under their associated tensorial $\delta_\Lambda$-transformations, there arises no problem of consistency of the field equations.

Since the dimensionality $D$ of space-time is arbitrary in our formulation, our results have innumerable applications. For example, we can even start in $D+E$-dimensions, and perform a separate dimensional reductions [15] into the final $D$-dimensions. In other words, we can replace the dimension $D$ in (3.6) by $D+E$, and perform the dimensional reduction [15] of the action $I_2$. Moreover, we can also construct some 'topological' terms in arbitrary $D$-dimensions. For example, when $D = 3r - 3$ for a given $r$ $(r \geq 3)$, we can have the $\delta_\Lambda$ and $\delta_\alpha$-invariant action

$$I_{\text{top}} \equiv \int d^D x \, \mathcal{L}_{\text{top}} = \int G_r{}^I G_{r-1}{}^{JK} G_{r-2}{}^{IJK}$$
$$= \int d^{3r-3}x \, \frac{1}{(3r-3)!} \epsilon^{\mu_1 \cdots \mu_r \nu_1 \cdots \nu_{r-1} \rho_1 \cdots \rho_{r-2}} G_{\mu_1 \cdots \mu_r}{}^I G_{\nu_1 \cdots \nu_{r-1}}{}^{JK} G_{\rho_1 \cdots \rho_{r-2}}{}^{IJK} \quad , \quad (4.1)$$

in addition to $I_2$ in (3.6). Notice that the most leading term in (4.1) is a total divergence, while other terms give non-zero trivial interactions. Needless to say, we can consider a separate multiplet starting with the rank $s (\neq r)$: $(C_s{}^I, C_{s-1}{}^{IJ}, \cdots, C_0{}^{I_1 \cdots I_{s+1}})$, and construct more variety of gauge invariant lagrangians out of these two multiplets of different ranks. With the adjoint index available for tensor fields of arbitrary ranks, our method opens up a new avenue for constructing interactions for non-Abelian tensor fields with arbitrary gauge symmetry.

We have dealt only with bosonic tensors in this paper. The next natural question is the application of similar mechanism to fermionic fields, or supersymmetric generalizations [11][12][16]. This work also suggests that there may well exist an alternative way to cure the conventional Velo-Zwanziger disease [9] associated with a spin 3/2 field, other than local supersymmetry [10].

This work is supported in part by NSF Grant # 0308246.

# References

[1] D.Z. Freedman and P.K. Townsend, Nucl. Phys. **B177** (1981) 282;
J. Thierry-Mieg and L. Baulieu, Nucl. Phys. **B228** (1983) 259.

[2] L. Baulieu and J. Thierry-Mieg, Phys. Lett. **144B** (1984) 221.

[3] L. Smolin, Phys. Lett. **137B** (1984) 379.

[4] M. Pernici, K. Pilch and P. van Nieuwenhuizen, Phys. Lett. **143B** (1984) 103.

[5] L. Baulieu, Nucl. Phys. **B266** (1986) 75; *'Field Anti-Field Duality, p-Form Gauge Fields and Topological Field Theories'*, `hep-th/9512026`.

[6] L. Baulieu, Phys. Lett. **184B** (1987) 23.

[7] S.N. Solodukhin, Nuovo Cim. **B108** (1993) 1275, `hep-th/9211046`.

[8] B. de Wit and H. Samtleben, Fort. für Phys. **53** (2005) 442; `hep-th/0501243`.

[9] G. Velo and D. Zwanziger, Phys. Rev. **D186** (1969) 1337, *ibid.* **D188** (1969) 2218.

[10] D.Z. Freedman, P. van Nieuwenhuizen and S. Ferrara, Phys. Rev. **D13** (1976) 3214;
S. Deser and B. Zumino, Phys. Lett. **62B** (1976) 335;
P. van Nieuwenhuizen, Phys. Rep. **68C** (1981) 189.

[11] S.J. Gates, Jr., M.T. Grisaru, M. Roček and W. Siegel, *'Superspace'* (Benjamin/Cummings, Reading, MA 1983).

[12] J. Wess and J. Bagger, *'Superspace and Supergravity'*, Princeton University Press (1992).

[13] P.K. Townsend, K. Pilch and P. van Nieuwenhuizen, Phys. Lett. **136B** (1984) 38.

[14] H. Nicolai and P.K. Townsend, Nucl. Phys. **B98** (1981) 257.

[15] N. Scherk and J.H. Schwarz, Nucl. Phys. **B153** (1979) 61.

[16] S.J. Gates, Jr., Nucl. Phys. **B184** (1981) 381.